\documentclass[5p,times]{elsarticle}

\usepackage{epsfig}
\usepackage{amssymb,amsmath,amsthm,graphics}
\usepackage{color}
\usepackage[colorlinks=true,linkcolor=blue,citecolor=blue]{hyperref}

\usepackage{lineno,hyperref}

\journal{Journal of Magnetism and Magnetic Materials}

\begin{document}
\begin{frontmatter}


\title{Ellipticity effects on diffusive magnon spin and heat transport in easy-plane ferromagnets}
\author[1]{Nicol\'as Vidal-Silva} 

\author[2]{Alejandro O. Leon\corref{mycorrespondingauthor}}
\cortext[mycorrespondingauthor]{Corresponding author}
\ead{aleonv@utem.cl}

\address[1]{ Departamento de Ciencias Físicas, Universidad de La Frontera, Temuco, Chile }
\address[2]{Departamento de F\'isica, Facultad de Ciencias  Universidad Tecnol\'ogica Metropolitana, Las Palmeras 3360, \~Nu\~noa 780-0003, Santiago, Chile}


\begin{abstract}
When a magnetic material hosts spin-wave excitations, or magnons, the
local magnetization can rotate in circular or elliptical orbits, the latter arising naturally in the presence of magnetic anisotropies transverse to the equilibrium magnetization. This article investigates the diffusive transport of elliptical magnons in easy-plane ferromagnets. Our analysis starts with the derivation of the magnon dispersion relation and magnon spin from the Landau-Lifshitz-Gilbert equation with a perpendicular magnetic anisotropy. Then, using the Boltzmann transport equation in the relaxation time approximation and perturbation analysis, the magnon-spin and magnon thermal conductivities are obtained, quantifying the magnon transport in the insulator. Our calculations demonstrate that, in both three- and two-dimensional systems, the effects of ellipticity on magnon transport coefficients result in an enhancement or a decrease, depending on whether magnets with a easy or hard perpendicular-to-plane axis are considered, respectively. On the other hand, our results predict an enhancement of the magnon heat transport for both easy- and hard-axis magnetic systems. Our study supports previous works on magnon ellipticity and makes a step towards clarifying its effect on magnon transport properties.
\end{abstract}


\end{frontmatter}

\section{Introduction}
While Landau and Lifshitz published their famous equation of motion for the magnetization in 1935~\cite{RefLL}, the research area of magnetization dynamics has continued to attract scientific attention, today focusing on smaller and faster devices~\cite{RefSerpico}. Beyond magnetic field-induced dynamics, the magnetization can be excited by currents~\cite{RefSTTReview,RefSTTSlon,RefSTTBer,RefSOT}, temperature gradients~\cite{RefSpinCaloritronics1,RefSpinCaloritronics2,RefSpinCaloritronics3}, surface acoustic waves~\cite{RefSAW1,RefSAW2}, and a voltage-controlled magnetic anisotropy~\cite{RefVCMA1,RefVCMAReview}. The dynamical responses of magnetic systems to such external drives are diverse, including auto-oscillations~\cite{RefLimitCycle1,RefLimitCycle2}, chaotic orbits~\cite{RefMagneticChaos1,RefChaoticPatterns}, and motion of magnetic textures, such as domain walls~\cite{RefDW} and skyrmions~\cite{RefSkyrmions}. 
Regarding magnetic materials, magnetic insulators such as yttrium iron garnet (chemical formula Y$_3$Fe$_5$O$_{12}$, abbreviated as YIG) are ideal for studying spin excitations, e.g., spin waves, due to the absence of Joule heating~\cite{RefInsulatronics}. Therefore, numerous efforts have been invested in characterizing and controlling their spin-wave spectrum~\cite{RefYIGSpectrum} and polarization~\cite{RefYIGMagnonPol}, energy renormalization~\cite{RefYIGEneergy}, magnetic anisotropy~\cite{RefREIG1}, and magnetic compensation points~\cite{RefREIG2}. While conduction electrons are the spin carriers in metals, \textit{magnons} (the quanta of spin waves) carry angular momentum and heat in magnetic insulators, providing the basis for the research field known as \textit{magnonics} or \textit{magnon spintronics}~\cite{RefMagnonics1,RefMagnonics2,RefMagnonics3}.
Magnons can be polarized circularly or elliptically, depending on how the local magnetization rotates around its equilibrium point. This article focuses on elliptical spin waves (magnons) and their diffusive transport in magnetic insulators.

In a \textit{metal}$\vert$\textit{magnetic insulator} bilayer, the flip of an electron spin at the interface transfers angular momentum to the insulator, i.e., a magnon is injected from the metal~\cite{RefInterfacialTransport}. At short distances magnons behave as a viscous fluid~\cite{RefHydrodynamics}, while at longer scales they can be transported over microns, as quantified by the magnon-spin diffusion length and the magnon-spin and magnon-heat conductivities~\cite{RefGreenFTransport,RefGeneralNonLocal, RefCornelissen}. Magnons promise many applications to information technologies, e.g., to cavity magnonics~\cite{RefMagnonAppl1} and magnon transistors~\cite{RefMagnonAppl2,RefMagnonAppl3}. Therefore, studying and engineering their conductivities continues attracting scientific interest. For example, the magnon-spin conductivity can be enhanced by orders of magnitude by reducing the thickness of YIG films~\cite{RefSuperLargeMSC} due to the dimensional crossover where only a few two-dimensional sub-bands contribute to transport. The underlying mechanisms to enhance the magnon-spin conductivity have been proposed in terms of softening of low-energy magnons in anisotropic ferromagnets~\cite{RefSoftMagnons}, compensating for the magnetic damping~\cite{RefAntidamping}, and reducing the magnetization~\cite{RefRedMagn}. The band contribution to magnon transport has been studied in Ref.~\cite{RefBandsContri}, finding that the lowest-energy acoustic band is the main contribution to the magnon-spin conductivity. Also, the control of magnon-spin transport has been achieved by electrically tuning the magnon density in YIG~\cite{RefMagnonControl} and in-plane anisotropies~\cite{RefLowFields}.
The topic of magnon-spin transport, particularly in the three- to two-dimensional crossover, has been recently studied in antiferromagnets~\cite{RefAFM}. Since thinner ferrimagnetic and ferromagnetic layers can exhibit more elliptical spin-wave polarizations due to their larger shape anisotropy, this raises the question of the role of ellipticity in magnon-spin transport. Reference~\cite{RefEllipticity2021} shows that in highly anisotropic systems a parasitic spin conductance can emerge~\cite{RefEllipticity2021}. The nonconservation of spin opens new transport channels~\cite{RefEllipticityandDamping}. However, to the best of our knowledge, the contribution of magnon ellipticity, via the dispersion relation and angular momentum per magnon, to magnon-spin transport remains unresolved for easy-plane magnets. 

In this work, we clarify the contribution of magnon ellipticity to diffusive magnon transport in easy-plane magnetic insulators. We model the system using the Landau-Lifshitz-Gilbert equation, from which the dispersion relation and spin of magnons are obtained. Then, using the Boltzmann transport equation in the relaxation time approximation, we calculate the magnon-spin and magnon thermal conductivities in both three- and two-dimensional systems and isolate the effect of ellipticity within a perturbative expansion. We find that increasing the easy-plane anisotropy parameter (i.e., the magnon ellipticity) modifies the magnonic dispersion relation and the spin per magnon, resulting in a reduction of the magnon-spin conductivity and an enhancement of the magnon thermal conductivity. These results indicate that the observed enhancements~\cite{RefSuperLargeMSC} of magnon transport in very thin magnets do not originate from ellipticity but from suppressed scattering in the quasi-2D limit, thereby clarifying the interpretation of experiments and guiding the design of magnonic devices.

The contribution of soft magnons to diffusive transport can be enhanced~\cite{RefSoftMagnons} near the critical points where the magnonic dispersion relation is non-analytic, which can be achieved at certain values of the applied magnetic field. Although Ref.~\cite{RefSoftMagnons} introduces the easy-plane in-plane-field geometry as its case (B), its transport analysis focuses on soft-magnon physics in another configuration, case (C). Here, we instead focus on diffusive magnon transport in the easy-plane in-plane field configuration and isolate the role of spin-wave ellipticity within a controlled perturbative expansion. In contrast to prior studies, the present article tracks ellipticity through both the modified (but fully analytic) dispersion and the spin per magnon, obtaining explicit ellipticity-induced corrections to the magnon-spin and magnon-heat conductivities in both three- and two-dimensional cases, thereby revealing the net ellipticity contribution.

\section{Elliptical magnons}
\label{SecEllipticManons}

Within a zero-temperature continuum description, the magnetization $\mathbf{M}=M_s\mathbf{m}$ has a constant norm $M_s$ and unit vector $\mathbf{m}=m_x\mathbf{e_x}+m_y\mathbf{e_y}+m_z\mathbf{e_z}$ that we decompose in the $\{x,y,z\}$ Cartesian representation of unit vectors $\{\mathbf{e_x},\mathbf{e_y},\mathbf{e_z}\}$.
The magnetization is subject to internal interactions, such as magnetocrystalline anisotropies and exchange, as well as an external magnetic field $\mathbf{B}=\mu_0M_s\mathbf{h}$, where $\mu_0$ is the magnetic permeability of free space. Next, we use a dimensionless position vector, $\mathbf{r}=\mathbf{r}_{\rm Ph}/l_{\rm ex}$, where $l_{\rm ex}$ is the exchange length of the magnet and $\mathbf{r}_{\rm Ph}$ is the position vector. The equilibrium magnetization configuration of a ferromagnet (or ferrimagnet) of volume $V_0$ minimizes the magnetic energy $E=\mu_0M_s^2V_0\int \epsilon \left[\mathbf{m}\right]d^3r$, with the following energy density $\epsilon$
\begin{equation}
\epsilon\left[\mathbf{m}\right]=-\mathbf{m}\cdot\mathbf{h}+\frac{\beta_zm_z^2}{2}+\frac{\parallel\boldsymbol{\nabla}\mathbf{m}\parallel^2}{2},
\label{EnergyDensity}
\end{equation}
where the first term is the Zeeman coupling to the magnetic field~\cite{RefSerpico}. The second term accounts for a uniaxial magnetic anisotropy, whose density is parametrized by the dimensionless constant $\beta_z$, quadratic in the magnetization component $m_z$. In the following, we concentrate on a planar magnet parallel to the $x-y$ plane. In this setup, $\beta_z$ becomes the perpendicular magnetic anisotropy constant that summarizes several effects, including the magnetodipolar interaction in the local shape anisotropy approximation, the interfacial perpendicular magnetic anisotropy, and bulk magnetocrystalline contributions. The magnetic layer has an easy-plane if $\beta_z>0$ and an easy-axis if $\beta_z<0$, favoring equilibrium configurations with $m_z\to0$ and $m_z\to\pm1$, respectively.
The nabla operator $\boldsymbol{\nabla}\equiv\mathbf{e_x}\partial_x+\mathbf{e_y}\partial_y+\mathbf{e_z}\partial_z$ is defined in terms of the partial derivatives with respect to the $j-$th spatial component, $\partial_j$. We use the notation $\parallel\boldsymbol{\nabla}\mathbf{m}\parallel^2=\vert\partial_x\mathbf{m}\vert^2+\vert\partial_y\mathbf{m}\vert^2+\vert\partial_z\mathbf{m}\vert^2$, where $\vert \mathbf{A}\vert$ stands for the Euclidean norm of the vector $\mathbf{A}\in\mathbb{R}^3$.
The temporal evolution of the magnetization is governed by the Landau-Lifshitz-Gilbert equation, which in its dimensionless form reads~\cite{RefSerpico}
\begin{equation}
\frac{\partial \mathbf{m}}{\partial t}=-\mathbf{m}\times\mathbf{h}_{\rm eff}+\alpha\mathbf{m}\times\frac{\partial \mathbf{m}}{\partial t}.
\label{EqLLG}
\end{equation}%
Here, the dimensionless time ($t$) depends on the time  coordinate, $t_{\rm Ph}$, as $t=\gamma \mu_0M_s t_{\rm Ph}$ with $\gamma$ being the gyromagnetic ratio. The first term of Eq.~(\ref{EqLLG}) produces counterclockwise precessions of $\mathbf{m}$ around the effective field $\mathbf{h}_{\rm eff}=-\left(\mu_0M_s^2V_0\right)^{-1}\delta E/\delta \mathbf{m}$,
\begin{equation}
\mathbf{h}_{\rm eff}=\mathbf{h}-\beta_zm_z\mathbf{e_z}+\nabla^2\mathbf{m},
\end{equation}
where $\nabla^2\mathbf{m}=\partial_{xx}\mathbf{m}+\partial_{yy}\mathbf{m}+\partial_{zz}\mathbf{m}$.
\begin{figure}[t!]
	\centering
	\includegraphics[width=8.5cm]{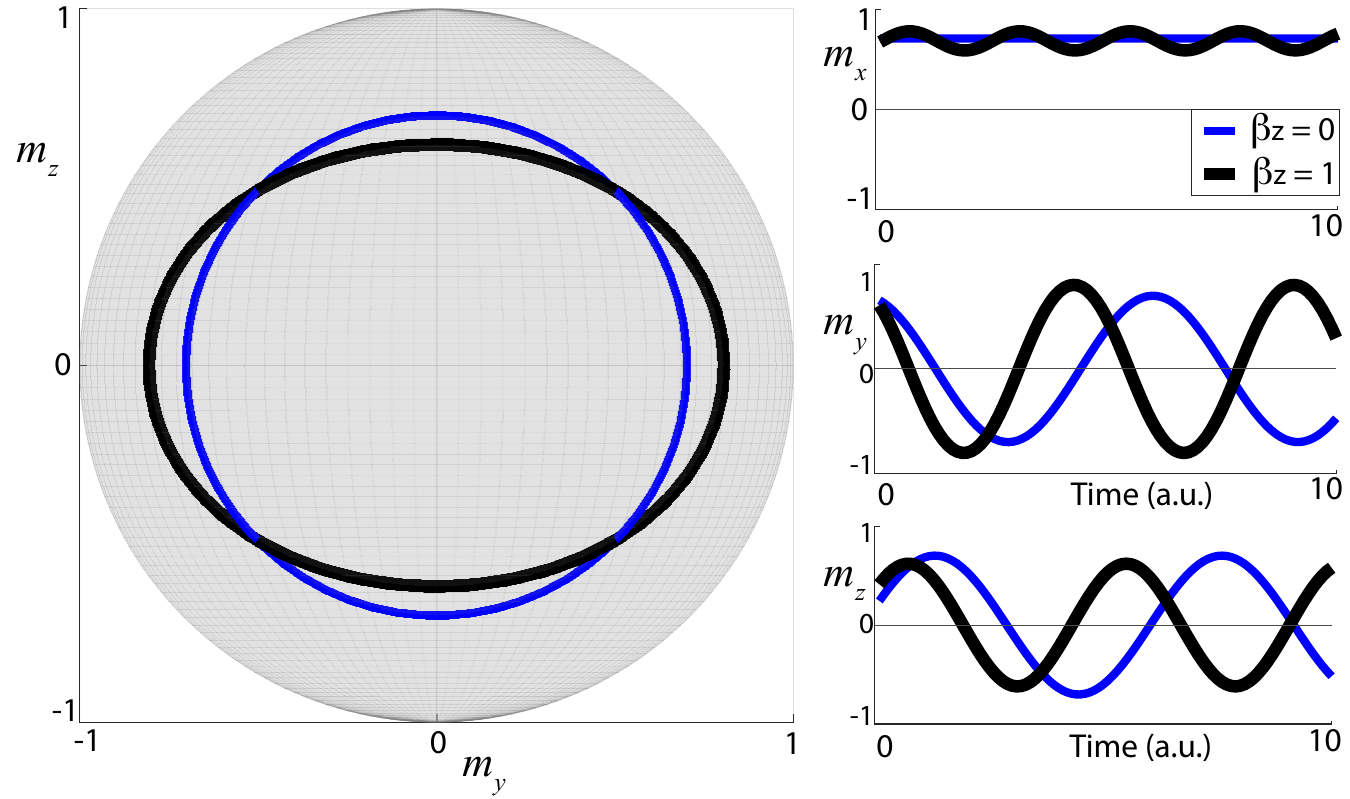}
	\caption{Comparison of a macrospin trajectory with (thicker black curve) and without (thinner blue curve) ellipticity, obtained for $\beta_z=1$ and $\beta_z=0$, respectively. Other parameters are $h=1$ and $\alpha=0.0001$. The left panel shows the magnetization trajectories of the magnetization vector $\left(m_x,m_y,m_z\right)$ parametrized by time $t$, over the unit sphere $\vert\mathbf{m}\vert=1$. As this plot illustrates, the ellipticity-free orbit (blue curve) is circular since the $m_y$ and $m_z$ components are equivalent energetically. On the other hand, the perpendicular magnetic anisotropy penalizes excursions with large $m_z$ [see Eq.~(\ref{EnergyDensity})], deforming the black orbit, that is, the anisotropy induces ellipticity in the oscillations. The right panel shows the Cartesian components of the magnetization. Note that for circular oscillations ($\beta_z=0$) the $m_x$ component is constant, but this component is not a conserved quantity for elliptical oscillations.
	The $m_y$ and $m_z$ plots reveal that the elliptical oscillations have a frequency shift, consistent with Eq.~(\ref{EqRelDispEll}), and that their amplitudes differ. Finally, note that for this very short time window, the small parameter $\alpha$ does not noticeably damp the oscillations.}
	\label{fig1}
\end{figure}
Let us choose the direction of the applied field along the $x-$axis, $\mathbf{h}=h\mathbf{e_x}$.
A rigidly moving magnetization, or macrospin, will be saturated along the $x-$axis by $\mathbf{h}$ and small deviations will rotate in the $y-z$ plane until reaching $\mathbf{m}=\mathbf{e_x}$ equilibrium. For $\beta_z=0$, this rotation is symmetric (circular) since it satisfies $m_y^2=m_z^2$, but it becomes asymmetric (elliptical) with finite $\beta_z$. This geometry is illustrated in Fig.~\ref{fig1} using a three-dimensional $\left(m_x,m_y,m_z\right)$ plot on the left panel and the individual magnetization components on the right panel. The data for this figure was obtained by numerically integrating of the LLG equation in the macrospin approximation. Note that for circular orbits ($\beta_z=0$), $m_x$ is constant and $m_y$ and $m_z$ display equivalent (phase-shifted) oscillations. However, perpendicular magnetic anisotropy ($\beta_z$) penalizes large $m_z$ values, resulting in an elliptical deformation of the orbit. This work aims to explore the consequences of such ellipticity for magnon transport.

\begin{figure}[b!]
	\centering
	\includegraphics[width=8.5cm]{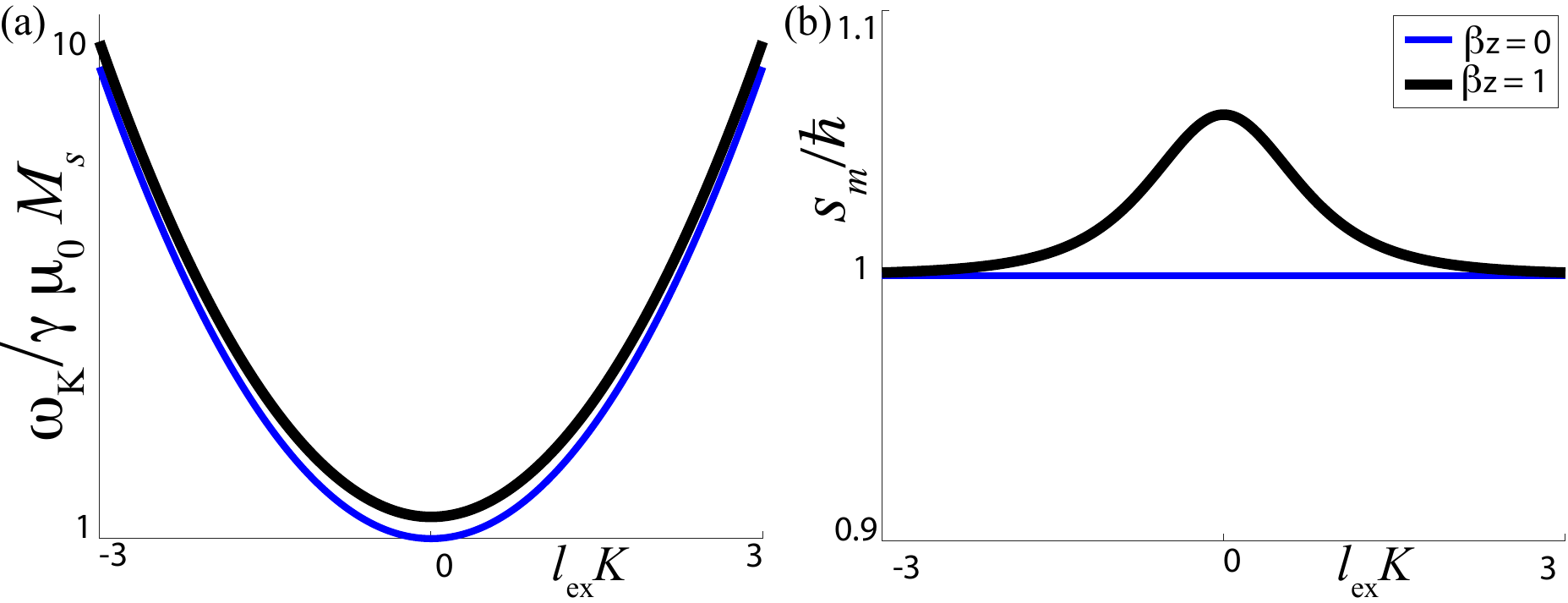}
	\caption{Magnon dispersion relation (a) and magnon spin (b) as a function of the wavenumber, normalized $l_{\rm ex}K$ for circular (thinner blue) and elliptic (thicker black) magnons, obtained for $\beta_z=0$ and $\beta_z=1$, respectively. As this plot illustrates, the major role of introducing an ellipticity parameter is to rise the energy ($\hbar\omega_\mathbf{k}$) and spin of low momentum magnons.}
	\label{fig2}
\end{figure}

After the straightforward calculations of~\ref{AppendixA}, and restorting the dimensions of the parameters, one obtains the magnon frequency
\begin{equation}
\omega_\mathbf{K}=\sqrt{\left(B+G+J_1K^2\right)^2-G^2},
\label{EqRelDispEll}
\end{equation}
where $J_1= \gamma\mu_0M_sl_{\rm ex}^2$ is the effective exchange parameter, $B= \gamma\mu_0M_sh$ is the external magnetic field, $G= \gamma\mu_0M_s\beta_z/2$ is the anisotropy field, and $\mathbf{K}=\mathbf{k}/l_{\rm ex}$ is the wave vector.
Figure~\ref{fig2}(a) shows the magnon dispersion relation for $\beta_z=0$ (corresponding to $G=0$) and $\beta_z=1$ ($G=\mu_0M_s/2$) as a function of the wavenumber. First, we notice that the relation is symmetric in momentum ($K\to-K$) and that the $G-$dependent contribution vanishes for large $\vert K \vert$ values since the frequency is dominated by the exchange term $\omega_\mathbf{k}\sim \gamma J_1K^2$. Additionally, note that without fields gap at $K=0$ closes, $\omega_{\mathbf{K}=0}\to0$ when $B\to0$.

In the presence of ellipticity, the spin per magnon $\left(s_m\right)$ differs from $\hbar$, and it is given by Hellmann-Feynman theorem~\cite{RefSoftMagnons},
\begin{align}
s_{\rm m}\equiv\hbar\frac{\partial\omega_\mathbf{K}}{\partial B}=
\frac{\hbar\left(B + G + J_1K^2\right)}{\sqrt{\left(B+G+J_1K^2\right)^2-G^2}},
\label{EqSmWithG}
\end{align}
which reduces to $s_m=\hbar$ for the circular case ($G=0$). Figure~\ref{fig2}(b) shows the magnon spin as a function of the wavenumber. The magnon spin has a maximum value $s_{\rm m}\left(K=0\right)=
\hbar\left(B + G \right)/\sqrt{B\left(B+2G \right)}$. However, this increment around $K=0$ is expected to play a minor role in the diffusive transport since magnons with higher momentum propagate.

The next section investigates how the diffusive spin transport is modified by the $G-$dependent terms in $\omega_\mathbf{K}$ and $s_{\rm m}$.

\section{Ellipticity in magnon transport}
\subsection{Transport of circular magnons}
Let us start by reviewing spin and heat transport of magnons with circular polarization ($G=0$). At thermodynamic equilibrium, magnons obey Planck's distribution
\begin{equation}
n_{\rm BE}\left(\frac{\hbar\omega_\mathbf{K}}{k_BT}\right)=\frac{1}{\exp\left(\frac{\hbar\omega_\mathbf{K}}{k_BT}\right)-1},
\end{equation}
which has zero magnon chemical potential, $\mu_{\rm m}=0$, and zero magnon heat accumulation, $\delta T_m=T_m-T=0$, i.e., the magnon temperature ($T_{\rm m}$) is equal to the ambient one ($T$) due to the nonconservation of magnons. However, under electric or thermal actuation, the magnon system exhibits the following quasi-equilibrium Bose-Einstein distribution due to the fast four-magnons scattering process~\cite{RefCornelissen}
\begin{equation}
f_{\rm qe}\left(\mathbf{r},\mathbf{K}\right)\equiv n_{\rm BE}\left(\frac{\hbar\omega_\mathbf{K}-\mu_{\rm m}\left(\mathbf{r}\right)}{k_BT_{\rm m}\left(\mathbf{r}\right)}\right),
\end{equation}
allowing one to define local values of $\mu_{\rm m}\left(\mathbf{r}\right)$ and $T_{\rm m}\left(\mathbf{r}\right)$.
Beyond the quasi-equilibrium statistics, the distribution function, $f$, satisfies the Boltzmann transport equation, which in the relaxation time approximation reads
\begin{equation}
\frac{df}{dt}\equiv\frac{\partial f}{\partial t}+\boldsymbol{\nabla}_\mathbf{K}f\cdot \frac{d\mathbf{K}}{dt}+\boldsymbol{\nabla}_\mathbf{r}f\cdot \mathbf{v}_G=-\frac{\delta f}{\tau},
\end{equation}
where the total time derivative of the distribution, $df/dt$, is written in terms of its partial time derivative $\partial f/\partial t$, the gradients in the reciprocal and direct spaces: $\boldsymbol{\nabla}_\mathbf{K}$ and $\boldsymbol{\nabla}_\mathbf{r}$, the time derivative of the wave number $d\mathbf{K}/dt$, and the group velocity $\mathbf{v}_G=\boldsymbol{\nabla}_\mathbf{K}\omega_\mathbf{K}$. Finally, $\tau$ is the relaxation time and $\delta f$ is the deviation from the quasi-equilibrium distribution $\delta f \left(\mathbf{r},\mathbf{K}\right)\equiv f\left(\mathbf{r},\mathbf{K}\right)-f_{\rm qe}\left(\mathbf{r},\mathbf{K}\right)$.
Since the YIG magnon interactions~\cite{RefYIGInt}~and lifetimes~\cite{RefMagnonLifeTime} are complex, treating the relaxation time as constant serves as a useful approximation that helps isolate the role of ellipticity in magnonic transport and allows us to obtain simple analytic expressions.

Considering a steady injection of angular momentum ($\partial f/\partial t=0$) and given the absence of magnonic Lorentz force ($d\mathbf{K}/dt=0$), the Boltzmann transport equation reduces to
\begin{equation}
\delta f=-\tau\boldsymbol{\nabla}_\mathbf{K}\omega_\mathbf{K}\cdot\boldsymbol{\nabla}_\mathbf{r}f_{\rm qe}\left(\mathbf{r},\mathbf{K}\right),
\label{EqDeltaF1}
\end{equation}
which can be used to derive the $D-$dimensional magnon transport, with $D=2,3$, including the magnon-spin $\left(\mathbf{j}_{\rm ms}\right)$ and magnon-heat $\left(\mathbf{j}_{\rm mQ}\right)$ currents
\begin{align}
\mathbf{j}_{\rm ms}&=e\int\frac{d^DK}{\left(2\pi\right)^D}\delta f\mathbf{v}_G\frac{s_{\rm m}}{\hbar},
\label{eq:jspin}\\
\mathbf{j}_{\rm mQ}&=\int\frac{d^DK}{\left(2\pi\right)^D}\delta f\mathbf{v}_G\hbar\omega_{\mathbf
K},\label{eq:jheat}
\end{align}%
 where the magnon-spin current $\mathbf{j}_{\rm ms}$ has units of charge-current density for later convenience.

To grasp the behavior of Eqs.~\eqref{eq:jspin} and \eqref{eq:jheat}, let us consider the high-temperature limit, $\left\vert\hbar\omega_\mathbf{K}-\mu_{\rm m}\left(\mathbf{r}\right)\right\vert \ll\left\vert k_BT_{\rm m}\left(\mathbf{r}\right)\right\vert\sim k_BT$. At leading order in the magnon-spin and heat accumulations,
\begin{equation}
\delta f=-\frac{\tau}{\left(\hbar\omega_\mathbf{K}\right)^2}
\boldsymbol{\nabla}_\mathbf{K}\omega_\mathbf{K}\cdot\left(k_BT\boldsymbol{\nabla}_\mathbf{r}\mu_{\rm m}+\hbar\omega_\mathbf{K}k_B\boldsymbol{\nabla}_\mathbf{r}T_{\rm m}\right),
\label{dfbyHTA}
\end{equation}
which is responsible for the transport when $\boldsymbol{\nabla}_\mathbf{r}\mu_{\rm m}\neq0$ or $\boldsymbol{\nabla}_\mathbf{r}T_{\rm m}\neq0$, that is, gradients on the magnon-spin and/or magnon-heat accumulations generate diffusive magnon transport.
For a parabolic dispersion relation, $\hbar\omega_\mathbf{K}=J_1K^2$ and transport along the $x-$th axis, $\boldsymbol{\nabla}_\mathbf{r}\mu_{\rm m}=\mathbf{e_x}\partial_x\mu_{\rm m}$ and $\boldsymbol{\nabla}_\mathbf{r}T_{\rm m}=\mathbf{e_x}\partial_xT_{\rm m}$, one gets
\begin{equation}
\delta f=-\frac{2\tau K_x}{J_1\hbar K^4}
\left(k_BT\partial_x\mu_{\rm m}+J_1K^2k_B\partial_xT_{\rm m}\right).
\label{eq:.deltaf}
\end{equation}
Within the linear response regime,
\begin{align}
\begin{pmatrix}
\mathbf{j}_{\text{ms}} \\
\mathbf{j}_{\text{mQ}}
\end{pmatrix}
=
\begin{pmatrix}
\sigma_\text{m} & S_{12} \\
S_{21} & \kappa_\text{m}
\end{pmatrix}
\begin{pmatrix}
-\boldsymbol{\nabla}_\mathbf{r}\mu_{\rm m}/e \\
-\boldsymbol{\nabla}_\mathbf{r}T_{\text{m}}
\end{pmatrix},
\label{eq:linearresponse}
\end{align}
where $\sigma_\text{m}$ and $\kappa_\text{m}$ are the magnon-spin and magnon thermal conductivities, while $S _{12}$ and $S_{21}$ are related to the Spin Seebeck and Spin Peltier coefficients \cite{RefCornelissen}. \\

For three-dimensional magnets ($D=3$), the spin current is obtained by replacing Eq. \eqref{eq:.deltaf} into Eq. \eqref{eq:jspin}
\begin{align}
\mathbf{j}_{\rm ms}=&-\frac{\tau ek_B\mathbf{e_x}}{\pi^2\hbar^2}\int_0^{K_\infty} dk\int_0^\pi d\tilde{\Omega}
\cdot\left(T\partial_x\mu_{\rm m}+J_1K^2\partial_xT_{\rm m}\right),\nonumber
\\
=&-\frac{2\tau ek_B\mathbf{e_x}}{3\pi^2\hbar^2}\int_0^{K_\infty} dk
\left(T\partial_x\mu_{\rm m}+J_1K^2\partial_xT_{\rm m}\right),
\\
=&-\frac{2\tau ek_B\mathbf{e_x}}{3\pi^2\hbar^2}
\left(TK_\infty\partial_x\mu_{\rm m}+\frac{J_1K_\infty^3}{3}\partial_xT_{\rm m}\right),
\label{eq:3Dspincurrent}
\end{align}
where $d\tilde{\Omega}\equiv d\theta_k\sin\left(\theta_k\right)\cos^2\left(\theta_k\right)$ and $K_\infty$ is a large momentum cutoff, indicating that large momentum magnons do not contribute much to transport. Comparing Eq.~(\ref{eq:3Dspincurrent}) with Eq.~(\ref{eq:linearresponse}), we get
\begin{align}
  \label{EqSigma3DCircular}  \sigma_{\rm m}^{(3D)}&=\frac{2\tau e^2k_B T}{3\pi^2\hbar^2}K_\infty.\\
    S_{12}^{(3D)}&=\frac{2\tau ek_B}{9\pi^2\hbar^2}J_1K_\infty^3.
\end{align}
Similarly, the magnon-heat current reads
\begin{align}
\nonumber \mathbf{j}_{\rm mQ}=&-\frac{2\tau}{3\pi^2\hbar^2}k_BJ_1\left( T\frac{K_{\infty}^3}{3}\partial_x\mu_{\text{m}}+J_1 \frac{K_{\infty}^5}{5}\partial_x T_{\text{m}}\right)
\end{align}
from which we identify the magnon-heat transport coefficients,
\begin{align}
 \kappa_{\rm m}^{(3D)}&=\frac{2\tau k_B J_1^2}{15\pi^2\hbar^2}K_{\infty}^5.\\
    S_{21}^{(3D)}&=\frac{2\tau ek_B TJ_1}{9\pi^2\hbar^2}K_\infty^3.
\end{align}

The previous results are consistent with Refs.~\cite{RefSuperLargeMSC,RefCornelissen}. We can obtain a closed value for the large momentum cutoff $K_{\infty}$. Indeed, by repeating the calculation for the magnon-spin current without introducing the high temperature approximation, i.e., using Eq.~(\ref{EqDeltaF1}) directly, yields
\begin{align}
\sigma_{\rm m}^{(3D)}=-e^2\frac{\partial }{\partial \mu_{\rm m}}\int\frac{d^3K}{\left(2\pi\right)^3}\delta f\frac{\partial \omega_\mathbf{K}}{\partial K_x}=\frac{2\tau e^2k_B T}{3\pi^2\hbar^2}\frac{C_0\sqrt{k_B T}}{\sqrt{2J_1}},
\label{EqSigma3DIsoFull}
\end{align}
where we used the numerical integration of the following integral
\begin{equation}
C_0\equiv\int_0^\infty du u^{3/2}\sinh^{-2}\left(u\right)\approx2.46.
\end{equation}
Equations~(\ref{EqSigma3DCircular}) and~(\ref{EqSigma3DIsoFull}) are the same after identifying
\begin{equation}
K_\infty=C_0\sqrt{\frac{k_B T}{2J_1}},
\label{eq:Klarge}
\end{equation}
which allows us to establish the $\sim T^{3/2}$ and $\sim T^{5/2}$ dependence of the spin and heat conductivities. It is important to point out that the large-momentum cutoff $K_{\infty}$ is introduced as a mathematical regulator to avoid nonphysical ultraviolet divergences introduced by the high-temperature approximation used in Eq.~(\ref{dfbyHTA}). Physically, $K_{\infty}$ should be interpreted as a limit of validity of the continuum transport theory, fixing the momentum scale beyond which the underlying approximations cease to be quantitatively accurate. At momenta comparable to (or larger than) this scale, nonparabolic corrections to the magnon dispersion may become relevant, and enhanced scattering processes (for instance, magnon-phonon interactions \cite{RefYIGInt}) are expected to play an increasing role, consistent with the ultraviolet-cutoff discussion in Ref.~\cite{RefSoftMagnons}. 
In this context, Ref.~\cite{RefJBarker} showed that, even at room temperature, the acoustic magnon band in YIG, well described by a parabolic dispersion, is highly populated and more relevant for spin transport. As a result, $K_{\infty}$ naturally emerges as a thermal cutoff whose value can be consistently estimated from Eq.~\eqref{eq:Klarge}. For YIG at temperatures up to $T=300$~K, the resulting large-momentum cutoff is of the order $K_{\infty}\sim 10^{9}\,\mathrm{m}^{-1}$.

\subsection{Ellipticity in the magnon-spin current}
To examine the spectral contribution of magnon ellipticity, as well as the $G-$dependent change in the magnon spin, in its transport, we start with the dispersion relation obtained in the previous section,
which reduces to the circular case when $G\to0$, i.e., $ \omega_\mathbf{K}=\gamma\left(B+J_1K^2\right)$. Note that the dispersion relation~(\ref{EqRelDispEll}) predicts uniform ($k=0$) instabilities if $B\leq0$. A clear example is the case of $B=-G$, which results in $\hbar\omega_\mathbf{K}=\sqrt{J_1^2K^4-G^2}$,
where the ground state is different from the one stated earlier, and therefore magnons with wave numbers smaller than $\left(G/J_1\right)^{1/2}$ are amplified since $\hbar\omega_\mathbf{K}$ is no longer a real-valued function. The magnon transport of the case $B>0$ and $G<0$, corresponding to a hard magnetic plane, has been studied previously in Ref.~\cite{RefSoftMagnons}.  Therefore, in the rest of this document, we consider an easy magnetic plane with an in-plane magnetic field $B,G\geq0$, for which the magnon transport, to the best of our knowledge, had not been studied yet.

For $D = 3$, the magnon-spin conductivity yields
\begin{align}
\nonumber\sigma_{\rm m}^{(3D)}&=-e^2\frac{\partial }{\partial \mu_{\rm m}}\int\frac{d^3K}{\left(2\pi\right)^3}\delta f\frac{\partial \omega_\mathbf{K}}{\partial K_x}\frac{s_{\rm m}}{\hbar},\\&
=\frac{2\tau e^2k_B TK_\infty}{3\pi^2\hbar^2}+\Delta \sigma_{\rm m}^{(3D)}(G),
\end{align}
where Eq.~(\ref{EqSmWithG}) was used for the spin of magnons. Expanding in Taylor series in $G$, with implicit small expansion parameter $G/\left(k_BT\right)$, one obtains in the high temperature limit, 
\begin{equation}
\Delta \sigma_{\rm m}^{(3D)}
\equiv
\frac{4  \tau e^2 k_B   T \left(K_0-K_\infty\right)G}{3 \pi^2 \hbar ^2J_1 K_0K_\infty},
\end{equation}
where $K_0$ is a large wavelength (small momentum) cutoff.
Defining the ratio between the $G-$dependent and $G-$independent contributions to $\sigma_{\rm m}^{(3D)}$,
\begin{align}
\delta_{\rm m}^{(3D)}\equiv&
\frac{\Delta \sigma_{\rm m}^{(3D)}}{\sigma_{\rm m}^{(3D)}\left(G\to0\right)}=-\frac{2 \vert K_0-K_\infty\vert}{J_1 K_0 K_\infty^2}G,
\label{EqDefoftheSigmaRatio}
\end{align}
one can appreciate a linear decrease of the spin conductivity with increasing $G$. Similarly, the ellipticity contribution to the $S_{12}$ coefficient is captured by
\begin{align}
        S_{12}^{(3D)}&=\frac{2\tau ek_B }{9\pi^2\hbar^2}J_1K_\infty^3 +\Delta S_{12}^{(3D)}(G),
\end{align}
with 
\begin{align}
    \Delta S_{12}^{(3D)}=\frac{2\tau ek_B(K_0-K_{\infty})}{3\pi^2\hbar^2}G.
\end{align}
By defining the ratio $\delta_{\rm S_{12}}^{(3D)}$
\begin{align}
    \delta_{\rm S_{12}}^{(3D)}\equiv\frac{\Delta S_{12}^{(3D)}}{S_{12}^{(3D)}(G\rightarrow 0)}=-\frac{3\vert K_0 - K_{\infty}\vert}{J_1K_{\infty}^3}G.
    \label{EqDefoftheS12Ratio}
\end{align}
For both Eqs.~\eqref{EqDefoftheSigmaRatio} and~\eqref{EqDefoftheS12Ratio}, our results reveal that ellipticity diminishes ($G>0$) or enhances ($G<0$) the magnon-spin transport for easy and hard magnetic planes, respectively. The dominant-order contribution of ellipticity is linear in the anisotropy constant.\\

Analogously to $K_{\infty}$, the small-momentum cutoff $K_0$ is introduced as a mathematical regulator to avoid nonphysical infrared divergences when magnons become nearly softened. As discussed in Ref.~\cite{RefSoftMagnons}, $K_0$ can be physically interpreted as the momentum scale below which the clean, continuum description of diffusive magnon transport ceases to be quantitatively accurate. In this regard, finite system size, residual in-plane anisotropies, or weak disorder can naturally account for the presence of such an infrared cutoff which effectively opens a small gap in the magnon spectrum and regularizes the soft-magnon contribution.
Accordingly, $K_0$ sets the lower bound of momenta for which the quadratic magnon dispersion and the associated diffusive transport framework remain applicable. Although long-wavelength magnons are highly populated at finite temperatures due to the Bose--Einstein distribution, diffusive transport depends not only on the occupation but also on the group velocity. Within our model, which assumes a constant magnon lifetime, magnons with strictly $K=0$ do not contribute to transport since their group velocity vanishes ($v_K \sim K$). However, the continuum description combined with the constant relaxation time approximation may otherwise lead to unphysical infrared divergences, which motivates the introduction of $K_0$ as a physically meaningful infrared regulator. In Ref.~\cite{RefSoftMagnons}, a residual in-plane anisotropy field of the order of $1$~mT is used to define the infrared cutoff in YIG, allowing one to estimate $K_0 \sim 10^{5}$--$10^{6}\,\mathrm{m}^{-1}$, satisfying $K_0\ll K_\infty$.

Repeating the analysis in two spatial dimensions $D=2$, we get the two-dimensional magnon-spin conductivity $\sigma_{\rm m}^{(2D)}$,
\begin{align}
\sigma_{\rm m}^{(2D)}\equiv&-e^2\frac{\partial }{\partial \mu_{\rm m}}\int\frac{d^2k}{\left(2\pi\right)^2}\delta f\frac{\partial \omega_\mathbf{K}}{\partial K_x}\frac{s_{\rm m}}{\hbar},\\=&
\frac{\tau e^2k_B T}{\pi\hbar^2} \ln\left(\frac{K_\infty}{K_0}\right)+\frac{ \tau e^2 k_BT (K_0^2-K_\infty^2) G }{\pi \hbar ^2J_1 K_0^2 K_\infty^2}.
\end{align}
The $G\to0$ limit has been discussed in Ref.~\cite{RefSuperLargeMSC}. For the limit $K_{\infty}\gg K_0$, the ellipticity-dependent increase of the magnon-spin conductivity is
\begin{equation}
\Delta\sigma_{\rm m}^{(2D)}\approx-\frac{  \tau e^2 k_BT G}{\pi \hbar ^2 J_1 K_0^2 }.
\end{equation}
As previously done, we define the conductivity ratio $\delta_{\rm m}^{(2D)}$,
\begin{equation}
\delta_{\rm m}^{(2D)}\equiv
\frac{\Delta\sigma_{\rm m}^{(2D)}}{\sigma_{\rm m}^{(2D)}\left(G\to0\right)}=-\frac{G }{J_1 K_0^2 \ln \left(\frac{K_\infty}{K_0}\right)}.
\end{equation}
On the other hand, the coefficient $S_{12}^{(2D)}$ is affected by ellipticity through
\begin{align}
 S_{12}^{(2D)} = \frac{e\tau J_1k_BK_{\infty}^2}{2\pi\hbar^2} - \frac{\tau ek_B}{\pi\hbar^2}\ln\left(\frac{K_{\infty}}{K_0}\right)G,
\end{align}
and the ratio $\delta_{S_{12}}^{(2D)}$ is

\begin{align}
\delta_{S_{12}}^{(2D)}\equiv
\frac{\Delta S_{12}^{(2D)}}{S_{12}^{(2D)}\left(G\to0\right)}&=-\frac{2 \ln\left(\frac{K_{\infty}}{K_0}\right)}{J_1K_{\infty}^2}G.
\end{align}
Similar to the $3D$ magnon spin current, in this case, ellipticity reduces ($G>0$) or enhances ($G<0$) the magnonic response for the spin transport. As a result, since the predicted variations in spin transport scale linearly with the anisotropy constant, one would expect a larger decrease in transport experiments for elongated samples with equilibrium magnetization perpendicular to the shape anisotropy easy axis.

\subsection{Ellipticity in the magnon heat current}
We now elucidate the effect of ellipticity on the transport coefficients associated with the magnon heat current in three- and two-dimensional systems. When including ellipticity in the 3D case, we obtain
\begin{align}
     \kappa_{\rm m}^{(3D)}&=\frac{2\tau k_B J_1^2}{15\pi^2\hbar^2}K_{\infty}^5 + \Delta\kappa_{\rm m}^{(3D)}(G).\\
    S_{21}^{(3D)}&=\frac{2\tau ek_B TJ_1}{9\pi^2\hbar^2}K_\infty^3 + \Delta S_{21}^{(3D)}(G).
\end{align}
where the ellipticity dependent coefficients read
\begin{align}
    \Delta\kappa_{\rm m}^{(3D)}=\frac{2\tau k_B K_{\infty}}{3\pi^2\hbar^2}G^2\\
    \Delta S_{21}^{(3D)}=-\frac{2\tau ek_BT K_{\infty}}{3\pi^2\hbar^2}G
\end{align}

To quantify the increase (decrease) of magnon heat transport by ellipticity, we define the ratios $\delta_{\rm Qm}^{(3D)}$ and  $\delta_{S_{21}}^{(3D)}$ as
\begin{align}
    \delta_{\rm Qm}^{(3D)}\equiv\frac{\Delta\kappa_{\rm m}^{(3D)}}{\kappa_{\rm m}^{(3D)}(G\rightarrow 0)}&=\frac{5G^2}{J_1^2K_{\infty}^4}\\
    \delta_{S_{21}}^{(3D)}\equiv\frac{\Delta S_{21}^{(3D)}}{S_{21}^{(3D)}(G\rightarrow 0)}&=-\frac{3G}{J_1K_{\infty}^2}.
\end{align}

Note that, differently from the spin transport coefficients, ellipticity enters quadratically into the correction in the magnonic heat conductivity. Therefore, our calculations predict that ellipticity always increases the magnon heat transport, regardless of whether the sample possesses an easy or hard magnetic axis. However, the coefficient $S_{21}$, related to the Spin Peltier effect, is affected in the same manner as the spin transport coefficients, implying an opposite response to the ellipticity for magnets with an easy or hard axis.
Next, for the 2D magnon heat transport we get 
\begin{align}
    \kappa_{\rm m}^{(2D)}&=\frac{J_1^2k_B\tau K_{\infty}^4}{4\pi\hbar^2}+\frac{\tau k_B \ln\left(\frac{K_{\infty}}{K_0}\right)}{\pi\hbar^2}G^2,\\
    S_{21}^{(2D)}&=\frac{e\tau J_1k_BTK_{\infty}^2}{2\pi\hbar^2}-\frac{e\tau k_B T  \ln\left(\frac{K_{\infty}}{K_0}\right)}{\pi\hbar^2}G,
\end{align}
from which we define the heat transport ratios
\begin{align}
    \delta_{\rm Qm}^{(2D)}\equiv\frac{\Delta\kappa_{\rm m}^{(2D)}}{\kappa_{\rm m}^{(2D)}(G\rightarrow 0)}&=\frac{4\ln\left(\frac{K_{\infty}}{K_0}\right)}{J_1^2K_{\infty}^4}G^2,\\
    \delta_{S_{21}}^{(2D)}\equiv\frac{\Delta S_{21}^{(2D)}}{S_{21}^{(2D)}(G\rightarrow 0)}&=-\frac{2 \ln\left(\frac{K_{\infty}}{K_0}\right)}{J_1 K_{\infty}^2}G.
\end{align}

Note that in both three- and two-dimensional systems, the presence of ellipticity has the same effect on magnon heat transport. As a result, ellipticity will increase the magnon thermal conductivity, while the heat transport mediated by a magnon chemical potential gradient is affected according to the sign of the magnetic anisotropy.\\

\section{Numerical estimates}

Here, we provide numerical estimates for the ellipticity-induced variations of both the magnon spin and the heat transport coefficients. We consider typical parameters for YIG at $T=300$ K, as it is commonly used in magnon-transport experiments. Consistent with the previous discussions, we take an ultraviolet cutoff $K_\infty\sim10^{9}\,\mathrm{m^{-1}}$ at room temperature and an infrared cutoff $K_0=10^{6}\,\mathrm{m^{-1}}$, compatible with the description of exchange magnons. In addition, we consider $J_1 = 5\times 10^{-17}$ Tm$^2$~\cite{RefSoftMagnons}. 
For bulk YIG samples in which the shape anisotropy is largely compensated, a residual anisotropy field 
of the order of a few mT \cite{RefSoftMagnons} can be taken as the main source of ellipticity. Thus, we adopt the anisotropy field $G/\gamma= 0.1$~mT, in which our perturbative expansion is valid. \\

Table~\ref{tab:yig1} summarizes the relative variations of the 2D and 3D spin and heat transport 
coefficients. For the 2D spin conductivity, we obtain 
$\delta_m^{(2D)}\approx-0.289$, implying a reduction of about $29\%$ with respect to the 
elipticity-free limit.
In three dimensions the correction  $\delta_m^{(3D)}\approx-(K_0/K_\infty)$ is strongly suppressed by the small ratio $K_0/K_\infty\sim10^{-3}$, leading to changes of about $0.4\%$.  For the thermal conductivity the ellipticity enters quadratically, yielding changes proportional to $(K_0/K_\infty)^{4}$  which are extremely small for $K_0\ll K_\infty$. Hence, ellipticity mainly affects 2D spin transport, while heat transport is enhanced only by a negligible amount within the diffusive 
regime of our model. Note that, although the correction to the 2D magnon spin conductivity appears sizable, it mainly originates from the enhanced infrared sensitivity characteristic of 2D transport integrals. Therefore, while quantitative values should be interpreted with caution, the qualitative trends predicted by our perturbative analysis are expected to remain robust under realistic experimental conditions.

\begin{table}[!]
\centering
\caption{Ellipticity-induced relative variations for $\mu_0H_{\rm ani}=0.2\,\mathrm{mT}$ (so $G=0.1$ mT), $K_0=10^{6}\,\mathrm{m^{-1}}$, $J_1 = 5\times 10^{-17}$ Tm$^2$ and $K_\infty=10^{9}\,\mathrm{m^{-1}}$ (i.e., $K_\infty/K_0=10^3$). The corrections are defined by $X(G)=X(0)[1+\delta]$.}
\label{tab:yig1}
\begin{tabular}{lcc}
\hline
\multicolumn{3}{c}{Two-dimensional transport}\\
\hline
Coefficient & $\delta$ & Variation \\
\hline
$\sigma_m^{(2D)}$ & $-2.8953\times10^{-1}$ & $-28.953\%$  \\
$S_{12}^{(2D)}$   & $-2.7631\times10^{-5}$ & $-0.0027631\%$ \\
$S_{21}^{(2D)}$   & $-2.7631\times10^{-5}$ & $-0.0027631\%$\\
$\kappa_m^{(2D)}$ & $1.10524\times10^{-10}$ & $1.10524\times10^{-8}\%$ \\
\hline
\multicolumn{3}{c}{Three dimensional transport}\\
\hline
$\sigma_m^{(3D)}$ & $-3.996\times10^{-3}$ & $-0.3996\%$  \\
$S_{12}^{(3D)}$   & $-5.994\times10^{-6}$ & $-0.0005994\%$ \\
$S_{21}^{(3D)}$   & $-5.994\times10^{-6}$ & $-0.0005994\%$ \\
$\kappa_m^{(3D)}$ & $2.0\times10^{-11}$ & $2.0\times10^{-9}\%$ \\
\hline
\end{tabular}
\end{table}

\section{Conclusions}
Classical spin waves, as well as their quantum counterpart, \textit{magnons}, can be circularly or elliptically polarized. In the elliptical case, the spin is not conserved, giving rise to a wide variety of phenomena. This work focused on the ellipticity-induced modifications to diffusive magnon transport by computing the magnon spin and thermal conductivities. For the easy magnetic plane and in-plane magnetic field configuration in three- and two-dimensional systems, we showed that the magnon-spin conductivity decreases with the anisotropy coefficient. This result supports that the enlargement of the magnon-spin conductivity in magnetic insulators with reduced dimensionality originates mainly from the suppression of scattering channels, i.e., in the $\tau$ parameter. 

On the other hand, the heat transport is differently affected by ellipticity. Indeed, our results reveal that in both two- and three-dimensional magnetic media, the magnon thermal conductivity is always enhanced by ellipticity, and this enhancement scales quadratically with the magnetic anisotropy constant, implying that the effect should be observed in both magnetic materials with easy and hard magnetic axes. \\

We emphasize that our results pertain to diffusive in-plane transport, where ellipticity reduces the magnon-spin conductivity but enhances the magnon thermal conductivity. By contrast, in tunneling spin valves, ellipticity and damping break spin conservation and open otherwise forbidden antiparallel channels~\cite{RefEllipticityandDamping}, whereas in lead-coupled geometries, strong anisotropy introduces a parasitic local spin resistance~\cite{RefEllipticity2021}. Our analysis, therefore, complements these studies by clarifying the diffusive case and establishing a minimal framework for studying magnon transport properties as a function of magnon ellipticity. Finally, note that the conclusions of this work do not rely on the specific values of the infrared and ultraviolet momentum cutoffs, $K_0$ and $K_{\infty}$. Instead, the relevant requirement is the existence of a momentum window $K_0 < K < K_{\infty}$ in which the continuum description of magnons and the diffusive transport framework remain valid. Within this window, the ellipticity-induced corrections to the magnon transport coefficients are controlled by the structure of the dispersion and by the effective spin per magnon. Indeed, we show that the ellipticity-induced effects in magnon transport depend on the size of such a window. For realistic parameters of YIG, the two cutoffs differ by several orders of magnitude, ensuring the robustness of our results with respect to microscopic details of the ultraviolet and infrared regularization.

It is important to note that in our framework ellipticity is not treated as an independent variable, but as a consequence of the same magnetic anisotropy that also modifies the magnon dispersion. Therefore, the changes in the transport coefficients originate from both the modification of the spin per magnon and the variations of the magnon spectrum, which affect group velocities and the corresponding phase space. Our definition of the “ellipticity effect’’ corresponds to the difference between the results for $G\neq0$ and the circular limit $G=0$ within the same Boltzmann description, without artificially separating these contributions.
 
\section*{Acknowledgments}
Alejandro O. Leon gratefully acknowledges financial support in Chile from FONDECYT Grant 11230120. Nicol\'as Vidal-Silva acknowledges FONDECYT Grant 1250364.

\appendix
\section{Mathematical discussion on the ellipticity of waves and magnon dispersion relation}
\label{AppendixA}

The LLG equation is a partial differential equation for the vector $\mathbf{m}$ that must be solved with the $\vert \mathbf{m}\vert=1$ constraint.  While there are several approaches to characterizing elliptical magnons, we consider here a transformation of the LLG equation to a general model of waves, which will allow us to discuss the ellipticity in a broader context. Indeed, using the following Holstein-Primakoff-type stereographic change of variables~\cite{RefSerpico}
\begin{equation}
A\left(\mathbf{r},t\right)=e^{i\frac{\pi}{4}}\frac{m_y+im_z}{1+m_x},
\label{EqAnsatzStereo}
\end{equation}
where the complex field $A\left(\mathbf{r},t\right)$ characterizes the deviations from the $m_x=1$ equilibrium. The amplitude $A=R e^{i\varphi_A}$ serves as a natural description of the waves envelope ($R\equiv\vert A\vert$) and phase ($\varphi_A$).
Replacing Ansatz~(\ref{EqAnsatzStereo}) in Eq.~(\ref{EqLLG}), one gets the following wave equation (see details in Ref.~\cite{RefEquivalenceAle1}),
\begin{equation}
\partial_t A=-i\nu A-i\beta_zA\vert A\vert^2-i\nabla^2 A-\mu A+\gamma A^\ast,
\label{EqPDNLS}
\end{equation}
where $A^\ast$ is the complex conjugate of $A$ and we have defined the following effective parameters: $\nu=-\left(h+\beta_z/2\right)$ is a frequency, $\mu=-\alpha\nu$ is a dissipation parameter, and the ellipticity parameter $\gamma=\beta_z/2$ accounts for the symmetry breaking in the complex plane of $A$, as shown later in this section. When only the term $-i\nu A$ is present in the right-hand side of Eq.~(\ref{EqPDNLS}), the solution in terms of an initial condition $A_0$ is $A=A_0\exp\left(-i\nu t\right)$. The role of the nonlinear term is to produce an amplitude-dependent shift in the oscillation frequency, $\nu\to\nu+\beta_z\vert A\vert^2$. The spatial coupling $-i\nabla^2 A$ accounts for the wave dispersion, while $-\mu A$ and $\gamma A^\ast$ are the reduce and modulate (typically increase) the wave energy. 
When $\mu=\gamma=0$, Eq.~(\ref{EqPDNLS}) is known as the \textit{nonlinear Schr\"odinger equation}~\cite{RefMikeska1978}, a conservative model that one can write as
\begin{align}
\left[\partial_t A\right]_{\mu=\gamma=0}&=-i\frac{\delta H}{\delta A^\ast},\\
H\left[A,A^\ast\right]&=\int d^3r\left[\frac{\nu\vert A\vert^2}{2}+\frac{\beta_z\vert A\vert^4}{4}+\frac{\vert \nabla A\vert^2}{2}\right],
\end{align}
where the functional $H$ plays the role of a Hamiltonian, such that $dH/dt=0$.  On the other hand, $\mu A$, with $\mu>0$, damps the oscillations driving the amplitude $A$ to its $A=0$ equilibrium. In this case, the dynamics is of a relaxation type, $dH/dt\leq0$. On the other hand, for finite ellipticity, $\gamma\neq0$, the relaxation structure is broken and Eq.~(\ref{EqPDNLS}) is referred to as \textit{parametrically driven, damped, nonlinear Schr\"odinger equation} (pdNLS), a so-called \textit{universal model} of waves describing the self-organization of systems forced at twice their natural frequency~\cite{RefpdNLSGeneral,RefParametricAle1,RefParametricAle2}. 
For example, the pdNLS equation governs the dynamics of surface waves in vibrated water~\cite{RefpdNLSFluids}, where the forcing parameter $\gamma$ is proportional to the amplitude of the channel vibration. Similarly, the pdNLS equation has been derived in a chain of pendula with time-dependent forcing~\cite{RefpdNLSMecha}, driven Kerr resonators~\cite{RefpdNLSOptics}, magnetic wires with oscillatory magnetic fields and Josephson junctions with time-dependent forcing~\cite{RefpdNLSSGandMagn}. Then, Eq.~(\ref{EqPDNLS}) shows that the magnon ellipticity is mathematically equivalent to a driving force in other physical systems, and this effective force generates the amplification of nonlinear waves and the emergence of solitary waves~\cite{RefpdNLSGeneral,RefParametricAle1,RefParametricAle2}.

In the present case, we obtain the pdNLS equation even if there is no temporal forcing because the anisotropy breaks the rotational symmetry in the $\left(m_y,m_z\right)$ plane as the temporal driving breaks the $t\to t+\Delta t$ in other systems~\cite{RefEquivalenceAle1}. This equivalence has been used to predict several magnetization structures~\cite{RefEquivalenceAle1}. In parametrically driven systems~\cite{RefpdNLSFluids,RefpdNLSMecha,RefpdNLSOptics,RefpdNLSSGandMagn}, the pdNLS equation is obtained from the equation of motion of the state variable, $\theta$, via a change of variable of the form $\theta\sim Ae^{i\Omega t}$. Therefore, the $\gamma A^\ast$ term is associated to the time-dependent forcing at frequency $\Omega$ (i.e., the force/torque that breaks the time-translation invariance). This parametric pumping can be made evident from Eq.~(\ref{EqPDNLS}) via the change of variables $A=e^{i\Omega t}C$,
\begin{equation}
\partial_tC=-i\left(\Omega+\nu\right)C-iC\vert C\vert^2-i\nabla^2 C-\mu C+\gamma e^{-2i\Omega t}C^\ast,
\end{equation}
where the last term is a periodic modulation of a parameter, i.e., a \textit{parametric driving.} Considering that the ellipticity term is equivalent to an injection of energy, it is natural to ask to what extent this symmetry-breaking term, $\gamma A^\ast$, can influence magnons and, in particular, their transport properties.

Let us concentrate on the small motions around the $A=0$ equilibrium for which $A\vert A\vert^2\approx0$. Furthermore, if $\gamma=0$, then $
\partial_t A=-i\nu A-i\nabla^2 A-\mu A$, whose general plane-wave solution reads
\begin{equation}
\left[A\left(\mathbf{r},t\right)\right]_{\gamma=0}=\sum_\mathbf{k}A_0\left(\mathbf{k}\right)e^{-\mu t+i\left(\omega_0 t+\mathbf{k}\cdot\mathbf{r}\right)},
\label{EqPlaneWaveSol}
\end{equation}
where the natural frequency is $\omega_0\left(k\right)=- \nu+k^2=h+\beta_z/2+k^2$, and $\mathbf{k}$ is a wavevector with modulus $k$. On the other hand, when $\gamma\neq0$, the formula~(\ref{EqPlaneWaveSol}) is not a solution of the wave equation, and a linear transformation has to be conducted to separate the $A$ and $A^\ast$ contributions to $\partial_tA$. After straightforward calculations, we get
\begin{align}
A\left(\mathbf{r},t\right)&=\gamma^{-1}\sum_\mathbf{k}\left(A_1+A_2\right)e^{-\mu t-i\left(\sqrt{\omega_0^2-\gamma^2}t+\mathbf{k}\cdot\mathbf{r}\right)},\label{EqParametricWaveSol}\\
A_1&=i B_0\left(\mathbf{k}\right)\gamma e^{2i\left(\sqrt{\omega_0^2-\gamma^2}t+\mathbf{k}\cdot\mathbf{r}\right)},\\
A_2&=B_0^\ast\left(\mathbf{k}\right)\left(\omega_0^2-\sqrt{\omega_0^2-\gamma^2}\right),
\end{align}
where $B_0\left(\mathbf{k}\right)$ is the complex amplitude of mode $\mathbf{k}$. In
the small $\gamma$ limit, $A\left(\mathbf{r},t\right)$ reduces to
\begin{equation}
A\left(\mathbf{r},t\right)\approx
\sum_\mathbf{k}\bar{B}_0^\ast\left(\mathbf{k}\right)e^{-\mu t-i\left(\omega_0t+\mathbf{k}\cdot\mathbf{r}\right)},
\end{equation}
where the amplitude $\bar{B}_0\left(\mathbf{k}\right)\equiv \gamma^{-1}B_0^\ast\left(\mathbf{k}\right)\omega_0\left(\omega_0-1\right)$ depends on the initial and boundary conditions. The corresponding eigen frequency $\left(\omega_0^2-\gamma^2\right)^{1/2}$ of elliptical magnons reads
\begin{equation}
\sqrt{\omega_0^2-\gamma^2}=\sqrt{\left(h+\beta_z/2+k^2\right)^2-\beta_z^2/4},
\end{equation}
which, after restoring the units, becomes Eq.~(\ref{EqRelDispEll}) of the main text.

\end{document}